\documentclass{article}

\usepackage{arxiv}

\usepackage[utf8]{inputenc} 
\usepackage[T1]{fontenc}    
\usepackage{hyperref}       
\usepackage{url}            
\usepackage{booktabs}       
\usepackage{amsfonts}       
\usepackage{nicefrac}       
\usepackage{microtype}      
\usepackage{lipsum}
\usepackage{graphicx}
\usepackage{epsfig}
\usepackage{subcaption}
\usepackage{xcolor}
\usepackage{pifont}
\usepackage{amsmath,amssymb} 
\usepackage{multirow}
\usepackage{upgreek}
\usepackage{arydshln}
\usepackage{tikz}
\usepackage{comment}
\usepackage{color}
\DeclareMathAlphabet{\pazocal}{OMS}{zplm}{m}{n}

\title{Effective Invertible Arbitrary Image Rescaling}

\author{
  Zhihong Pan, Baopu Li \\
  Baidu Research (USA)\\
   \\
   \And
 Dongliang He, Wenhao Wu, Errui Ding \\
  Department of Computer Vision Technology (VIS), Baidu Inc.\\
  \\
}

\begin{document}
\newcommand{\cmark}{\ding{51}}%
\newcommand{\xmark}{\ding{55}}%
\newcommand{\red}[1]{\textcolor{red}{#1}}
\newcommand{\blue}[1]{\textcolor{blue}{#1}}
\newcommand{\green}[1]{\textcolor{green}{#1}}
\newcommand{\teal}[1]{\textcolor{teal}{#1}}
\newcommand{\orange}[1]{\textcolor{orange}{#1}}
\newcommand{\wt}[1]{\textcolor{white}{#1}}
\newcommand{\etal}{\textit{et al.}}

\maketitle

\begin{abstract}
Great successes have been achieved using deep learning techniques for image super-resolution (SR) with fixed scales. To increase its real world applicability, numerous models have also been proposed to restore SR images with arbitrary scale factors, including asymmetric ones where images are resized to different scales along horizontal and vertical directions. Though most models are only optimized for the unidirectional upscaling task while assuming a predefined downscaling kernel for low-resolution (LR) inputs, recent models based on Invertible Neural Networks (INN) are able to increase upscaling accuracy significantly by optimizing the downscaling and upscaling cycle jointly. However, limited by the INN architecture, it is constrained to fixed integer scale factors and requires one model for each scale. Without increasing model complexity, a simple and effective invertible arbitrary rescaling network (IARN) is proposed to achieve arbitrary image rescaling by training only one model in this work. Using innovative components like position-aware scale encoding and preemptive channel splitting, the network is optimized to convert the non-invertible rescaling cycle to an effectively invertible process. It is shown to achieve a state-of-the-art (SOTA) performance in bidirectional arbitrary rescaling without compromising perceptual quality in LR outputs. It is also demonstrated to perform well on tests with asymmetric scales using the same network architecture.
\end{abstract}

\vspace{-5pt}
\section{Introduction}
\label{sec:intro}

\begin{figure*}[t]
 \begin{center}
     \includegraphics[width=\linewidth]{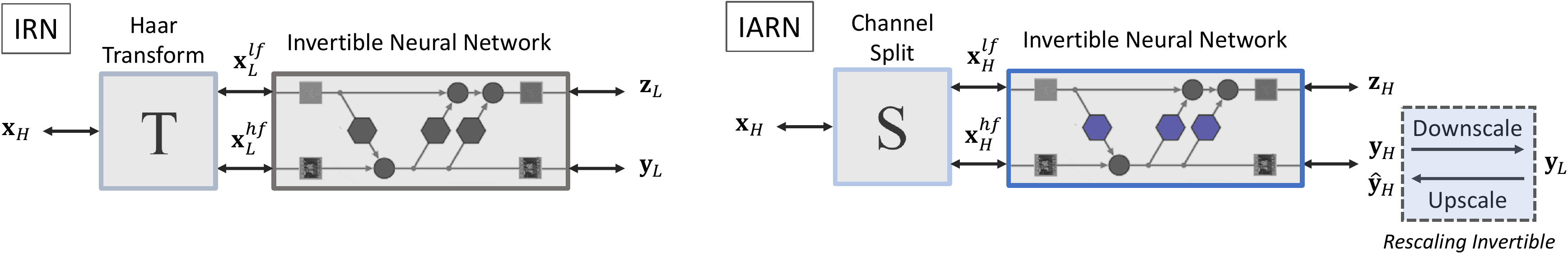}
 \end{center}
 \vspace{-6pt}
 \caption{Comparison between IRN~\cite{xiao_eccv_2020} and the proposed IARN. $\mathbf{x}_H$, $\mathbf{y}_L$ and $\mathbf{z}_{H, L}$ denote input HR image, output LR image and the latent variables.  Note that subscripts $H, L$ refer to high- and low-resolution respectively, and superscripts $hf, lf$ refer to high and low frequency channels respectively.  The main differences in IARN include channel-splitting in high-resolution in place of Haar transform, the additional near-invertible rescaling between $\mathbf{y}_H$ and $\mathbf{y}_L$, and enhancement in transformation blocks inside the INN backbone.}
 \vspace{-9pt}
 \label{fig:iarn}
 \end{figure*}

Recent deep learning based image super-resolution (SR) methods have advanced the
performance of image upscaling significantly but they are often limited
to fixed integer scale factors and pre-determined downscaling degradation kernels.
To work in real world applications where an image is commonly rescaled to arbitrary sizes,
additional image resizing is often needed, which leads to degradation
in both performance and efficiency. 
Lately there are growing interests in SR models that support arbitrary scale factors and great successes
have been achieved in recent works \cite{chen_cvpr_2021,hu_cvpr_2019,wang_iccv_2021}.
However, they are only optimized for the unidirectional upscaling process
with the LR inputs either synthesized from a predefined downscaling kernel or in its native resolution.
Considering the potential mutual benefits between downscaling and the inverse upscaling,
some image rescaling models~\cite{kim_eccv_2018,sun_tip_2020,xiao_eccv_2020}
are developed to optimize these two processes jointly
and significant improvements in upscaling accuracy are achieved comparing to unidirectional SR models of the same
scale factors.  The state-of-the-art (SOTA) performance for such bidirectional image rescaling is set by the
invertible rescaling net (IRN) as proposed by Xiao \etal~\cite{xiao_eccv_2020}.
As shown in Fig.~\ref{fig:iarn}, it is able to achieve the best performance so far since both the Haar transformation
and the invertible neural network (INN)~\cite{ardizzone_iclr_2018} backbone are invertible processes, {and}
its forward and invertible backward operations can model the downscaling and inverse upscaling cycle
naturally.  Denoting the forward downscaling process as
$(\mathbf{y}_L, \mathbf{z}_L) = f(\mathbf{x}_H)$, the HR image can be fully restored as
$\mathbf{x}_H = f^{-1}(\mathbf{y}_L, \mathbf{z}_L)$ if the latent variable $\mathbf{z}_L$ is preserved.
When the network is optimized to store as much information as allowed in $\mathbf{y}_L$ and convert $\mathbf{z}_L$ as input-independent random variables,
the optimal HR output $\hat{\mathbf{x}}_H$ can be restored as $f^{-1}(\mathbf{y}_L, \hat{\mathbf{z}}_L)$ where $\hat{\mathbf{z}}_L$ is randomly sampled
with minimized loss in restoration accuracy.  However, limited by the nature of INN architecture that the number of pixels must be equivalent
when differences in resolution are accounted for (LR features have more channels), the applicable scale factors are limited to integers.


To overcome the above issues, we propose a new invertible arbitrary rescaling network (IARN) that is able to implement bidirectional arbitrary rescaling using
similar INN backbone and it is able to train one model for a range of arbitrary scales, including asymmetric ones.
As shown in Fig.~\ref{fig:iarn}, it replaces the Haar wavelet transformation blocks in IRN with a preemptive channel-splitting step denoted as
 \vspace{-4pt}
 \begin{equation}
 \begin{split}
 \mathbf{x}^{lf}_H & = s(\mathbf{x}_H) \\
 \mathbf{x}^{hf}_H & = \mathbf{x}_H-\mathbf{x}^{lf}_H
 \end{split}
 \end{equation}
\noindent where subscript $H$ is used to specify that all images are preserved in original high-resolution.  Note that this step is intrinsically
invertible as $\mathbf{x}_H = \mathbf{x}^{lf}_H+\mathbf{x}^{hf}_H$ for the inverse direction. The feature transformation step using
similar INN backbone, which is also invertible, is denoted as
 \vspace{-4pt}
 \begin{equation}
 \begin{split}
 (\mathbf{y}_H, \mathbf{z}_H) = f(\mathbf{x}^{lf}_H, \mathbf{x}^{hf}_H)
 \end{split}
 \label{eq:inn}
 \end{equation}
The downscaling and inverse upscaling are included as the last step and it consists of two separate
processes as
 \vspace{-4pt}
 \begin{equation}
 \begin{split}
 \mathbf{y}_L & = d(\mathbf{y}_H) \\
 \hat{\mathbf{y}_H} & = u(\mathbf{y}_L)
 \end{split}
 \label{eq:du}
 \end{equation}
To optimize the network for invertible arbitrary rescaling, in addition to the similar objectives to maximize information saved
in $\mathbf{y}_H$ and make $\mathbf{z}_H$ signal independent, one key challenge is to make the last step invertible.
If it satisfies that $\hat{\mathbf{y}_H} = \mathbf{y}_H$, the whole network is invertible.
While this bidirectional process in Equation~\ref{eq:du} is not invertible in general, for a given scale factor and specific downscaling function $d(\cdot)$ and upscaling $u(\cdot)$,
like nearest neighbour (NN) or bicubic interpolation, we can find a set of images which are invertible to this rescaling process
and our goal is to transform $\mathbf{y}_H$ as one of these rescaling invertible images so the full process is effectively invertible.
To help the transformation of $\mathbf{y}_H$ and make it approaching invertibility faster, a
new channel-splitting step $s(\cdot)$ that converts $\mathbf{x}^{lf}_H$ as rescaling invertible preemptively is proposed
and it is shown to be very effective.
Lastly, as the aforementioned rescaling-invertible feature is scale-dependent, an innovative position-aware scale encoding is proposed
as additional inputs of the network to make the network capable of handling large variations in arbitrary scale factors, including asymmetric ones.
This component is not illustrated in Fig.~\ref{fig:iarn} for simplicity reasons but will be discussed in details later.  In summary, the main contributions of our work include:
\begin{itemize}
\setlength\itemsep{0.01em}
\item[$\bullet$] The first to develop an invertible neural network for bidirectional arbitrary image rescaling, which sets a new SOTA performance
considering both generated LR and restored HR.

\item[$\bullet$] A preemptive channel-splitting step, which separates the rescaling-invertible component from the input HR image, 
is advanced to make the learning of invertible arbitrary rescaling more effective.

\item[$\bullet$] A position-aware scale encoding, which is independent of input image size
and compatible with asymmetric rescaling, is proposed to further boost model performance in a large range of arbitrary scales.

\end{itemize}

\section{Related Works}
\label{sec:rwork}

\noindent\textbf{{Arbitrary Scale Super-Resolution.}}
Single image super-resolution, as a form of image rescaling with a fixed integer scale factor
like $\times 2$ and $\times 4$, has been studied extensively.  For the last few years, deep learning based methods like
~\cite{dong_eccv_2014,kim_cvpr_2016_2,lim_cvprw_2017,zhang_cvpr_2018,zhang_eccv_2018}
have brought great successes in this field,
but these methods commonly train one model for each scale factor.
More recent models proposed by Lim \etal~\cite{lim_cvprw_2017} and Li \etal~\cite{li_eccv_2018}
are capable of training one model for multiple scaling factors but only limited to integer ones.
Inspired by the weight prediction techniques in meta-learning~\cite{lemke_air_2015},
a single Meta-SR model was proposed by Hu \etal~\cite{hu_cvpr_2019} to solve image rescaling
of arbitrary scale factors,
by predicting weights of convolutional layers for arbitrary scale factors within certain range.
Alternatively, Behjati \etal~\cite{behjati_wacv_2021} proposed the OverNet to generate over-upscaled
maps from which HR images of arbitrary scales can be recovered using optimal downsampling.  
In the latest ArbSR, Wang \etal~\cite{wang_iccv_2021} proposed a plug-in module that can
optimize existing SR models for arbitrary asymmetric rescaling,
where scale factors along horizontal and vertical directions could be different.
While these methods are often limited to a maximum scale factor like $\times 4$
to maintain high performance, Chen \etal~\cite{chen_cvpr_2021}
proposed recently to use learned pixel representation features to replace pixel value features in previous methods.  Using
the innovative local implicit image function (LIIF), this model can extrapolate well to out-of-distribution large scales that are not seen in training.
Different from the above models which are optimized for upscaling reconstruction only, we consider the bidirectional arbitrary rescaling
of downscaling and upscaling as one process in this work.

\vspace{3pt} \noindent\textbf{{Bidirectional Image Rescaling.}}
To take advantage of the potential
mutual benefits between downscaling and the inverse upscaling,
Kim \etal~\cite{kim_eccv_2018} proposed an auto-encoder framework
to jointly train image downscaling and upscaling together.
Similarly, Sun \etal~\cite{sun_tip_2020} suggested a new image downscaling method
using a content adaptive-resampler, which can be jointly trained with
any existing differentiable upscaling (SR) models.
More recently, Xiao \etal~\cite{xiao_eccv_2020} advanced an invertible rescaling net (IRN)
that has achieved SOTA performance for learning based bidirectional image rescaling.
Based on the invertible neural network (INN)~\cite{ardizzone_iclr_2018},
IRN learns to convert HR input to LR
output and an auxiliary latent variable $z$.  By mapping $z$ to a
case-agnostic normal distribution during training, inverse image upscaling is implemented
by randomly sampling $\hat{z}$ from the normal distribution without need of the case specific $z$.
While the above bidirectional image rescaling methods are limited to a fixed integer scale factor
like $\times 4$,  Pan \etal~\cite{pan_cvpr_2022} proposed the BAIRNet
as the first to solve bidirectional arbitrary rescaling by utilizing local implicit image functions
for both downscaling and upscaling with better overall cross-scale performance over IRN.
Most recently, Xing \etal~\cite{xing_arxiv_2022} proposed an
encoder-decoder network (AIDN) to tackle the same challenge with
consistent improvements over both IRN and BAIRNet.
Instead of using separate encoder and decoder to model image downscaling and upscaling respectively,
we propose here to model the downscaling and upscaling processes as forward
and backward operations of one INN backbone.  While very similar, their performances in the upscaling task are
consistently lower than ours with very few exceptions in certain scales.  Additionally,
LR outputs from our IARN have better perceptual quality comparing to AIDN in both blind or non-blind image
quality assessments.  As it is shown in \cite{li_arxiv_2021} that there is a performance trade-off between the generated LR and restored HR,
our IARN is obviously the superior one comparing to AIDN when both downscaling and upscaling tasks are considered.

\section{Proposed Method}
\label{sec:method}

 \begin{figure*}[t]
 \begin{center}
     \includegraphics[width=0.9\linewidth]{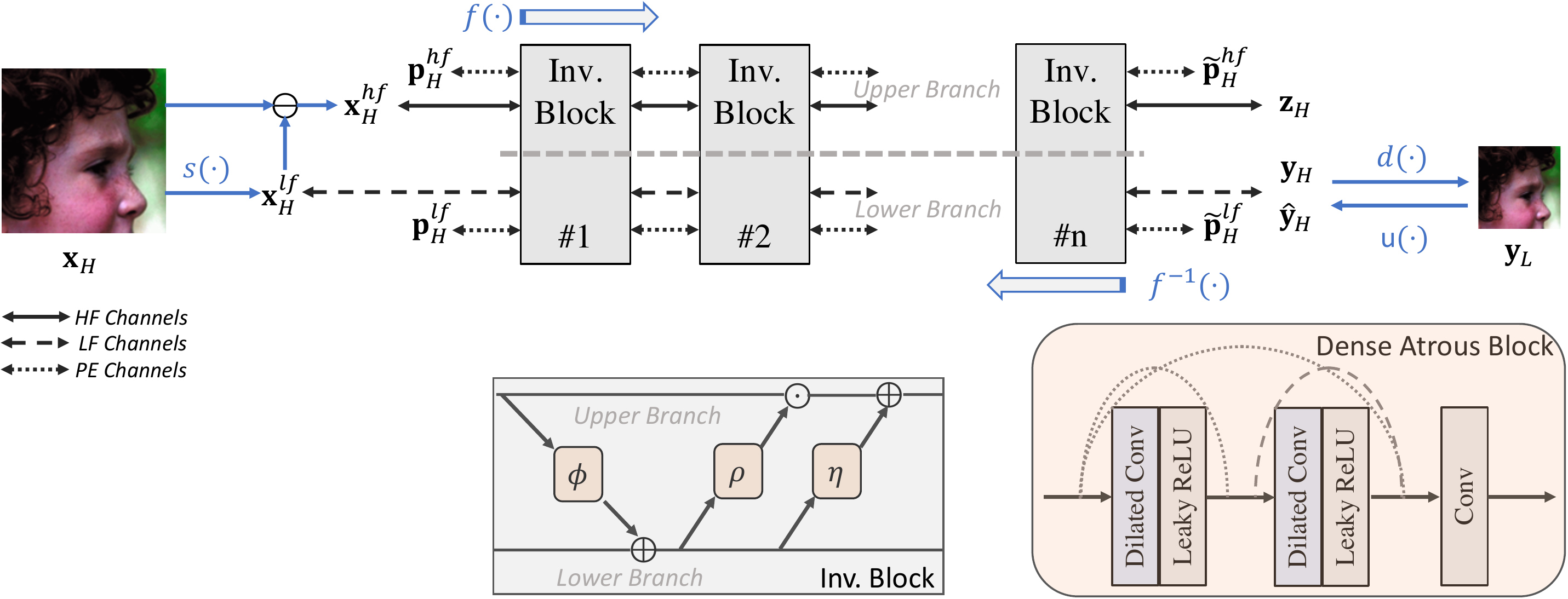}
 \end{center}
 \vspace{-3pt}
 \caption{Network architecture of the proposed IARN, including the full process of preemptive channeling splitting $s(\cdot)$, the feature transformation INN backbone and the rescaling-invertible module
 consists of downscaling $d(\cdot)$ and upscaling $u(\cdot)$, and the enhanced dense atrous block used for transformation blocks $\phi, \rho, \eta$.}
 \vspace{-12pt}
 \label{fig:net}
 \end{figure*}
\subsection{Network Architecture}

Elaborated from the general pipeline shown in Fig.~\ref{fig:iarn},
the detailed architecture of the proposed IARN is illustrated in Fig.~\ref{fig:net}.
For the forward downscaling process to generate LR output $\mathbf{y}_L$ from HR input $\mathbf{x}_H$,
the overall process is summarized in the following four steps:
 \begin{equation}
 \begin{split}
 \mathbf{x}^{lf}_H & = s(\mathbf{x}_H) \\
 \mathbf{x}^{hf}_H & = \mathbf{x}_H-\mathbf{x}^{lf}_H \\
 \mathbf{y}_H, \mathbf{z}_H, \tilde{\mathbf{p}}^{lf}_H, \tilde{\mathbf{p}}^{hf}_H & = f(\mathbf{x}^{lf}_H, \mathbf{x}^{hf}_H, \mathbf{p}^{lf}_H, \mathbf{p}^{hf}_H) \\
 \mathbf{y}_L & = d(\mathbf{y}_H)
 \end{split}
 \label{eq:ds}
 \end{equation}
\noindent where $s(\cdot)$ is the preemptive channel splitting function detailed below, $f(\cdot)$ is the forward function of the INN backbone
and $d(\cdot)$ is a downscaling function like nearest neighbour or bicubic interpolation.
For variables, $\mathbf{x}$ is the input in forward direction, $\mathbf{y}$ is the output, and $\mathbf{p}$ and $\tilde{\mathbf{p}}$ are the position-aware scale encoding at the input and output side respectively.  Details about this position-aware scale encoding will be introduced in later subsection.
For subscripts and superscripts, $H$ and $L$ means high- and low-resolutions, while $hf$ and $lf$ represents high and low frequency components respectively.
Similarly the inverse upscaling process consists of the following three steps:
 \begin{equation}
 \begin{split}
 \mathbf{y}_H & = u(\mathbf{y}_L) \\
 \mathbf{x}^{lf}_H, \mathbf{x}^{hf}_H, \mathbf{p}^{lf}_H, \mathbf{p}^{hf}_H & = f^{-1}(\mathbf{y}_H, \mathbf{z}_H, \tilde{\mathbf{p}}^{lf}_H, \tilde{\mathbf{p}}^{hf}_H) \\
 \mathbf{x}_H & = \mathbf{x}^{lf}_H+\mathbf{x}^{hf}_H
 \end{split}
 \label{eq:us}
 \end{equation}
\noindent where $u(\cdot)$ is an upscaling function and $f^{-1}(\cdot)$ is the inverse function of $f(\cdot)$.  

The process presented in Equation~\ref{eq:us} is an ideal inverse process, where $\mathbf{y}_H$ is perfectly restored after the downscaling and upscaling
cycle, and $\mathbf{z}_H, \tilde{\mathbf{p}}^{lf}_H, \tilde{\mathbf{p}}^{hf}_H$ are all preserved.  This ideal situation
does not exist in real application where only $\mathbf{y}_L$ is saved and $\hat{\mathbf{y}}_H \neq \mathbf{y}_H$.
Similar to the previous studies, a generated $\hat{\mathbf{z}}_H$ is
used in place of $\mathbf{z}_H$.  For scale encoding, as they are utilized to carry
position and scale factor information for each pixel for both forward and backward directions, and they are independent
of low and high-frequency branches, one general $\mathbf{p}$ is used for $\mathbf{p}^{lf}_H$ and $\mathbf{p}^{hf}_H$ during forward process
and the same for $\tilde{\mathbf{p}}^{lf}_H$ and $\tilde{\mathbf{p}}^{hf}_H$ during backward upscaling.
As a result, the restored HR output $\hat{\mathbf{x}}_H$ is calculated as
 \begin{equation}
 \begin{split}
 \hat{\mathbf{y}}_H & = u(\mathbf{y}_L) \\
 \hat{\mathbf{x}}^{lf}_H, \hat{\mathbf{x}}^{hf}_H & = f^{-1}(\hat{\mathbf{y}}_H, \hat{\mathbf{z}}_H, \mathbf{p}_H) \\
 \hat{\mathbf{x}}_H & = \hat{\mathbf{x}}^{lf}_H+\hat{\mathbf{x}}^{hf}_H
 \end{split}
 \label{eq:xhat}
 \end{equation}

While the primary goal is to minimize the restoration error between $\hat{\mathbf{x}}_H$ and $\mathbf{x}_H$, the overall
loss used to train the network includes multiple losses for different objectives as shown below
\begin{equation}
\vspace{0pt}
     L = \lambda_1 L_{r}+\lambda_2 L_{g}+\lambda_3 L_{d}+\lambda_4 L_{i}.
\vspace{0pt}
\label{eq:loss}
\end{equation}
Here $L_{r}$ is the $L1$ reconstruction loss for upscaled HR output $\hat{\mathbf{x}}_H$ and
$L_{g}$ is the $L2$ guidance loss for downscaled LR output $\mathbf{y}_L$ in reference to
a downsampled LR reference $\overline{\mathbf{y}}_L$ using bicubic interpolation.
For $L_{d}$, it is similar to the distribution regulation loss of latent variable $\mathbf{z}$ as in IRN~\cite{xiao_eccv_2020}.
However, with the newly introduced position-aware scale encoding, in place of the $L2$ loss between $\mathbf{z}_H$
and $\mathbf{z}^0_H$, it is calculated between $(\mathbf{z}_H, \tilde{\mathbf{p}}^{lf}_H, \tilde{\mathbf{p}}^{hf}_H)$
and reference $(\mathbf{z}^0_H, \mathbf{p}_H, \mathbf{p}_H)$. Here $\mathbf{z}^0_H$ is a zero tensor, following
the practice in IRN for the surrogate distribution
matching loss $L^\prime_{distr}$.
The last term $L_{i}$ is a novel rescaling-invertible loss that aims to make the full process effectively invertible.
As discussed earlier, the proposed process is only partially invertible since the downscaling
and upscaling cycle is not invertible.  In general cases, $\hat{\mathbf{y}}_H = u(d(\mathbf{y}_H)) \neq \mathbf{y}_H$.
However, given specific pair of $d(\cdot)$ and $u(\cdot)$, there are certain subset of rescaling invertible
images $\mathbf{y}^i_H$ satisfies $\mathbf{y}^i_H = u(d(\mathbf{y}^i_H))$.  For example, as explained later in
Section~\ref{sec:pcs}, for NN resampling, we can generate a rescaling-invertible
$\mathbf{x}^{lf}_H$ from any input image $\mathbf{x}_H$.
The rescaling-invertible loss $L_{i}$ is introduced here as the $L2$ loss between $\hat{\mathbf{y}}_H$ and $\mathbf{y}_H$.
When the loss is zero, then $\mathbf{y}_H$ is rescaling invertible.

Note that while the layers of Invertible Blocks (InvBlock) in our INN backbone is similar to the ones in IRN~\cite{wang_eccv_2018},
there are some key differences.  First,  position-aware scale encoding is introduced to both the lower low-frequency (LF)
and upper high-frequency (HF) branches to enable the network adaptive to arbitrary scales.
Secondly, for transformation functions between the upper and lower branches, including
$\phi(.)$, $\rho(.)$ and $\eta(.)$, the original densely connected convolutional block in IRN is enhanced with 
dilated convolutions where the dilation varies from 1 to $l$, which is the number of layers in each block.
$l$ is simplified as 2 here in Fig.~\ref{fig:net}.  This enhanced Dense Atrous Block is introduced to increase
effective receptive fields of the network with mixed dilation larger or equal to one.
As the process in the INN backbone is applied
to original high-resolution images, the receptive fields of our IARN is the same as IRN in pixel units,
but would be smaller in terms of image regions given the resolution difference if the network structure and depth remains
the same.  Using a mixture of dilated convolutions in place of the original ones, it can increase the receptive
field without changes in the number of model parameters and model complexity.

\subsection{Preemptive Channel Splitting}
\label{sec:pcs}

As seen in the Fig.~\ref{fig:net}, the HR output $\mathbf{y}_H$ is located in the lower branch of INN backbone
and matches the low-frequency HR input $\mathbf{x}^{lf}_H$ in channel orders.  While the transformation functions
in the INN backbone can transfer features between lower and upper branches quite efficiently, it may be beneficial
to pre-process the inputs so that $\mathbf{x}^{lf}_H$ is close to $\mathbf{y}_H$ even before feeding into the INN backbone, making
the learning task of transformation functions even more efficient.  As discussed above, it is ideal for $\mathbf{y}_H$
to be rescaling invertible for a given pair of resampling operation $d(\cdot)$ and $u(\cdot)$.  Here a preemptive channel splitting
function is proposed as $s(\cdot) = u(d(\cdot))$.  As pointed out in BAIRNet~\cite{pan_cvpr_2022},
in the case of integer scale factors,
$s(\cdot)$ would be an idempotent operation when using bilinear downscaling and nearest-neighbour upscaling.
That is, for any image $\mathbf{x}$, $s(s(\mathbf{x})) = s(\mathbf{x})$.  In other
words, $s(\mathbf{x})$ is rescaling-invertible.  For arbitrary scales here, $s(\mathbf{x})$ is only
rescaling invertible when using nearest-neighbour for both downscaling and upscaling.  Thus the preemptive
channel splitting, which splits the input image to two branches, is proposed here as
\begin{equation}
 \begin{split}
     \mathbf{x}^{lf}_H & = u_{N}(d_{N}(\mathbf{x}_H)) \\
     \mathbf{x}^{hf}_H & = \mathbf{x}_H-\mathbf{x}^{lf}_H \\
 \end{split}
\end{equation}
where $N$ refers to nearest-neighbour resampling.  Note that the pair of $d(\cdot)$ and $u(\cdot)$
must be the same for both channel splitting in the front and rescaling at the end.
Experiments are also conducted to demonstrate the advantage of nearest-neighbour over other choices of resampling
choices.

\subsection{Position-Aware Scale Encoding}

Both the preemptive channel splitting and rescaling invertible loss help the network to transform
$\mathbf{y}_H$ to induce minimum losses during downscaling and rescaling steps and lead to optimal
restoration of $\hat{\mathbf{x}}_H$ consequently.  As this transformation is sensitive to scale
factors, to enable the model robust to a large range of arbitrary scales, scale information are necessary during
the transformation between lower and upper branches.  Additionally, depending on the scale factor and rescaling
method, the position of the pixel inside the image is also important to determine operations applied to the pixel itself.
Using nearest-neighbour as an example, the resampling coefficients vary by pixel locations
because for some pixels, their nearest neighbours are located to the top-left while  some others located to the bottom-right.
To account for these factors, a position-aware scale encoding $\mathbf{p}$ is defined as
$(\mathbf{s}_h, \mathbf{s}_v, \mathbf{d}_h, \mathbf{d}_v)$.  Here $\mathbf{s}_h$ and $\mathbf{s}_v$
are the scale factors along horizontal and vertical directions respectively, to accommodate
optimal asymmetric scales.  For $\mathbf{d}_h$ and $\mathbf{d}_v$, they are the horizontal and vertical distances
from the input pixel to the closest resampled pixel to its bottom-right direction and can be calculated as below
\begin{equation}
 \begin{split}
 \mathbf{d}_h(i, j) & = \min_{i^{\prime}, i^{\prime} s^{\prime}_h - i s_h \geq 0} i^{\prime} s^{\prime}_h - i s_h \\
 \mathbf{d}_v(i, j) & = \min_{j^{\prime}, j^{\prime} s^{\prime}_v - j s_v \geq 0} j^{\prime} s^{\prime}_v - j s_v
 \end{split}
 \vspace{-1pt}
\label{eq:dist}
\end{equation}
where $(i, j)$ and $(i^{\prime}, j^{\prime})$ are indices of the input pixel and resample pixel respectively,
$s_h$ and $s_v$ are pixel sizes along horizontal and vertical axes for input image, and $s^{\prime}_h$ and $s^{\prime}_v$ are pixel sizes of resample image.
This distance is specifically selected to be aware of relative pixel positions but independent of image sizes.
The latter is necessary as the input image size is limited during training but larger out-of-distribution
image sizes are common for inferences during testing.
Experiments conducted later also demonstrate that the position-aware scale encoding is greatly beneficial
to network performance and it is the best to include it in both lower and upper branches.

\begin{table*}[ht!]
	\centering
	\footnotesize
	\setlength{\tabcolsep}{5pt}
\vspace{3pt}
	\caption{Quantitative comparisons of SOTA SR and rescaling methods with the best two results highlighted in \red{red} and \blue{blue} respectively (IARN$^\dagger$ and IARN$^\ddagger$ are used for intra-model comparisons only and not ranked, and methods in \textbf{bold} means requirement of multiple models and additional interpolations to conduct tests in arbitrary scales).}
\vspace{3pt}
	\begin{tabular}{cccccccc} 
		\hline
		{Method$\wt{^{^a}}$} & {Scale} & {Set5} & {Set14} & {BSD100} & {Urban100} & {Manga109} & {DIV2K}\\
		\hline \hline
		\textbf{RCAN}$\wt{^{^a}}$~\cite{zhang_eccv_2018}& {1.5} & 40.97/0.9767 & 37.05/0.9578 & 35.59/0.9516 & 35.93/0.9660 & 42.33/0.9889 & 38.47/0.9701\\
		Meta-SR~\cite{hu_cvpr_2019}& {1.5} & 41.47/0.9785 & 37.52/0.9601 & 35.86/0.9543 & 36.91/0.9696 & 43.17/0.9904 & 38.88/0.9718\\
		LIIF~\cite{chen_cvpr_2021}& {1.5} & 41.23/0.9774 & 37.37/0.9591 & 35.76/0.9536 & 36.70/0.9684 & 42.84/0.9894 & 38.82/0.9717\\
		ArbSR~\cite{wang_iccv_2021}\vspace{1pt} & {1.5} & 41.47/0.9786 & 37.51/0.9603 & 35.86/0.9547 & 36.92/0.9697 & 43.12/0.9904 & 38.84/0.9719 \\
		\hline
		\textbf{CAR}~\cite{sun_tip_2020}$\wt{^{^a}}$ & {1.5} & 40.50/0.9763 & 37.08/0.9596 & 35.72/0.9535 & 34.70/0.9635 & 40.90/0.9881 & 37.93/0.9683\\
		\textbf{IRN}~\cite{xiao_eccv_2020}& {1.5} & 43.55/0.9891 & 39.52/0.9795 & 39.28/0.9833 & 36.52/0.9811 & 42.64/0.9936 & 40.18/0.9838\\
	    BAIRNet~\cite{pan_cvpr_2022} & {1.5} & {47.13/0.9849} & {43.12/0.9760} & {46.63/0.9959} & {44.01/0.9946} & \blue{45.49/0.9948} & {44.99/0.9920}\\
	    AIDN~\cite{xing_arxiv_2022} & {1.5} & \blue{50.61/0.9961} & \blue{46.70/0.9920} & \blue{49.82/0.9983} & \blue{46.26/0.9967} & {-\protect\footnotemark} & \red{47.01/0.9953}\\
	    {IARN}$\wt{A}$ \vspace{1pt} & {1.5} & \red{51.02/0.9968} & \red{47.25/0.9938} & \red{50.91/0.9986} & \red{47.58/0.9975} & \red{48.58/0.9975} & \blue{46.74/0.9949} \\
		\hline \hline
		\textbf{RCAN}~\cite{zhang_eccv_2018}& {2.5} & 36.05/0.9436 & 31.69/0.8815 & 30.47/0.8508 & 30.42/0.8990 & 36.59/0.9634 & 32.72/0.9079\\
		Meta-SR~\cite{hu_cvpr_2019}& {2.5} & 36.18/0.9441 & 31.90/0.8814 & 30.47/0.8508 & 30.57/0.9003 & 36.55/0.9639 & 32.77/0.9086\\
		LIIF~\cite{chen_cvpr_2021}& {2.5} & 35.98/0.9434 & 31.64/0.8813 & 30.45/0.8510 & 30.42/0.8992 & 36.39/0.9630 & 32.78/0.9091\\
		ArbSR~\cite{wang_iccv_2021}\vspace{1pt} & {2.5} & 36.21/0.9448 & 31.99/0.8830 & 30.51/0.8536 & 30.68/0.9027 & 36.67/0.9646 & 32.77/0.9093 \\
		\hline
		\textbf{CAR}~\cite{sun_tip_2020}$\wt{^{^a}}$ & {2.5} & 37.33/0.9548 & 33.78/0.9169 & 32.53/0.9020 & 32.19/0.9301 & 37.63/0.9717 & 34.32/0.9310\\
		\textbf{IRN}~\cite{xiao_eccv_2020}& {2.5} & 39.78/0.9742 & 36.39/0.9553 & 35.56/0.9540 & 33.99/0.9589 & 39.33/0.9836 & 36.60/0.9607\\
	    BAIRNet~\cite{pan_cvpr_2022} & {2.5} & {40.11/0.9664} & {36.62/0.9469} & {36.29/0.9563} & {36.62/0.9679} & \blue{40.26/0.9830} & {37.46/0.9627}\\
	    AIDN~\cite{xing_arxiv_2022} & {2.5} & \blue{40.77/0.9750} & \blue{37.62/0.9588} & \blue{36.65/0.9593} & \red{37.10/0.9710} & {-\footnotemark[\value{footnote}]} & \blue{37.88/0.9659}\\
	    {IARN}$\wt{A}$ \vspace{1pt} & {2.5}  & \red{40.93/0.9756} & \red{37.78/0.9598} & \red{36.81/0.9607} & \blue{36.95/0.9703} & \red{41.38/0.9862} & \red{37.92/0.9662} \\
		\hline \hline
		\textbf{RCAN}~\cite{zhang_eccv_2018}& {3.5} & 33.47/0.9138 & 29.24/0.8141 & 28.42/0.7731 & 27.61/0.8348 & 32.74/0.9328 & 30.13/0.8511\\
		Meta-SR~\cite{hu_cvpr_2019}& {3.5} & 33.59/0.9146 & 29.60/0.8140 & 28.42/0.7728 & 27.71/0.8356 & 32.75/0.9337 & 30.18/0.8524\\
		LIIF~\cite{chen_cvpr_2021}& {3.5} & 33.41/0.9133 & 29.20/0.8131 & 28.39/0.7714 & 27.60/0.8334 & 32.60/0.9324 & 30.16/0.8517\\
		ArbSR~\cite{wang_iccv_2021} \vspace{1pt}& {3.5} & 33.63/0.9149 & 29.58/0.8147 & 28.41/0.7744 & 27.69/0.8360 & 32.84/0.9339 & 30.14/0.8518 \\
		\hline
		\textbf{CAR}~\cite{sun_tip_2020}$\wt{^{^a}}$ & {3.5} & 34.98/0.9303 & 31.38/0.8643 & 30.14/0.8326 & 29.97/0.8871 & 35.00/0.9507 & 31.88/0.8865\\
		\textbf{IRN}~\cite{xiao_eccv_2020}& {3.5} & 37.12/\blue{0.9546} & {33.65/0.9196} & 32.54/\blue{0.9047} & 31.84/0.9277 & 36.86/0.9690 & 33.84/0.9281\\
	    BAIRNet~\cite{pan_cvpr_2022} & {3.5} & 36.85/0.9472 & 32.97/0.9074 & 32.36/0.8986 & {32.71/0.9338} & {36.98/0.9671} & {33.87/0.9266}\\
	    AIDN~\cite{xing_arxiv_2022} & {3.5} & \blue{37.25}/0.9538 & \blue{33.87/0.9197} & \blue{32.73}/0.9032 & \blue{33.20}/\red{0.9372} & {-\footnotemark[\value{footnote}]} & \blue{34.19/0.9292}\\
	    {IARN}$\wt{A}$ \vspace{1pt} & {3.5}  & \red{37.44/0.9547} & \red{34.04/0.9218} & \red{32.90/0.9058} & \red{33.27}/\blue{0.9371} & \red{37.91/0.9705} & \red{34.33/0.9308} \\
	    \hdashline
	    IARN$^{\dagger\wt{^A}}$ & {3.5} & 36.82/0.9538 & 33.66/0.9194 & 32.69/0.9032 & 33.02/0.9350 & 37.72/0.9697 & 34.19/0.9292 \\
	    IARN$^\ddagger\wt{A}$ & {3.5} & 37.27/0.9537 & 33.74/0.9175 & 32.68/0.9019 & 32.81/0.9326 & 37.59/0.9690 & 34.09/0.9282 \\
		\hline	\end{tabular}
\label{tab:main}
\vspace{1pt}
\end{table*}

\section{Experiments}
\label{sec:exp}

\subsection{Data and Settings}
\label{sec:set}
For fair comparisons with recent relevant works like IRN and BAIRNet, the DIV2K~\cite{agustsson_ntire_2017}
training set is used for baseline training.  Another dataset, Flickr2K~\cite{timofte_ntire_2017}, is also
included for training the final model.
For quantitative evaluation, we use HR images from six commonly used datasets for comprehensive comparison,
including Set5 \cite{bevilacqua_bmvc_2012}, Set14~\cite{zeyde_iccs_2010},
BSD100~\cite{martin_iccv_2001}, Urban100~\cite{huang_cvpr_2015}, Manga109~\cite{huang_cvpr_2015},
and the DIV2K validation set.  
Following previous practices, we take the peak noise-signal ratio (PSNR) and
SSIM~\cite{wang_tip_2004} on the luminance channel for all test sets, with the exception of DIV2K which uses average of RGB channels.

To accommodate the large range of arbitrary scale factors, a total of 20 layer of invertible blocks are included
in the INN backbone, less than the total of 24 used in IRN$\times2$ and IRN$\times4$.
For the Dense Atrous Block, 4 layers of dilated convolution are included with dilation setting as 1 to 4 consecutively.
Mini-batches of 16 $144 \times 144$ patches are initially used, each with a
random scale sampled from an uniform distribution of $\pazocal{U}(1, 4)$.
It is upgraded to 24 $192 \times 192$ patches for the final model when trained using both DIV2K and Flickr2K.
There are two stages of training for the final models, each with $250k$ iterations where
the learning rate is reduced by half after each $50k$ iterations.
Settings of the two stages are the same except for the starting learning rate, which is
$2 \times 10^{-4}$ for the first and changed to $1 \times 10^{-4}$ when resuming the second stage training.
The weights of losses are set at empirically at 1, 16 and 2 for $L_{r}$, $L_{g}$ and $L_{i}$ respectively.
As we have found out in experiments, and similar to the findings by Li \etal~\cite{li_arxiv_2021},
setting $\hat{\mathbf{z}}$ as zero for training and inference achieves equivalent or better performance
comparing to randomly sampled $\hat{\mathbf{z}}$.  As a result, $L_{d}$ is set as zero for our experiments as it has little impact on learning.

\subsection{Arbitrary Rescaling Performance}
\vspace{-2pt}
To assess the performance of our proposed method for arbitrary rescaling in symmetric scales, we compare restoration
quality of rescaled HR images in a set of arbitrary scales and the results are included
in Table~\ref{tab:main}.  Similar to BAIRNet,
models trained for fixed integer scales
are evaluated for arbitrary scales
using additional bicubic interpolations to keep the resolution of LR images fed into
respective models identical across different methods.
For models capable of arbitrary scales on their own, most of them are optimized for upscaling only
so they are only assessed here for HR outputs.
Other than our IARN, BAIRNet is the only one that is trained for bidirectional arbitrary image rescaling.

\begin{figure*}[t!]
\captionsetup[subfigure]{font=footnotesize, labelformat=empty}
\begin{center}
  \begin{subfigure}[b]{0.12\textwidth}
    \centering
      \includegraphics[width=\textwidth, interpolate=false]{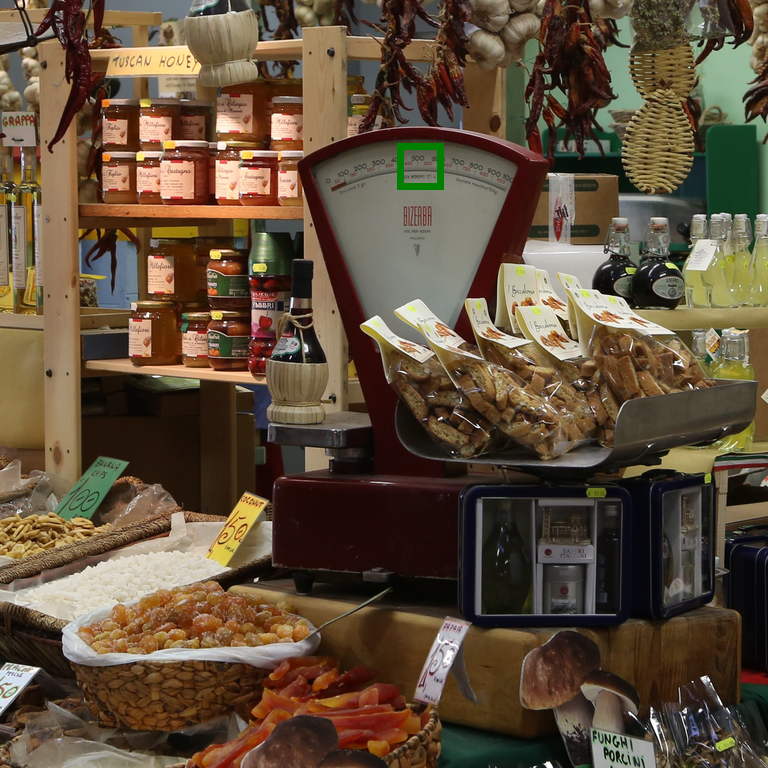}
  \end{subfigure} \hspace*{-0.45em}
  \begin{subfigure}[b]{0.12\textwidth}
    \centering
      \includegraphics[width=\textwidth, interpolate=false]{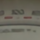}
  \end{subfigure} \hspace*{-0.45em}
  \begin{subfigure}[b]{0.12\textwidth}
    \centering
      \includegraphics[width=\textwidth, interpolate=false]{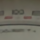}
  \end{subfigure} \hspace*{-0.45em}
  \begin{subfigure}[b]{0.12\textwidth}
    \centering
      \includegraphics[width=\textwidth, interpolate=false]{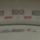}
  \end{subfigure} \hspace*{-0.45em}
  \begin{subfigure}[b]{0.12\textwidth}
    \centering
      \includegraphics[width=\textwidth, interpolate=false]{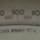}
  \end{subfigure} \hspace*{-0.45em}
  \begin{subfigure}[b]{0.12\textwidth}
    \centering
      \includegraphics[width=\textwidth, interpolate=false]{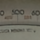}
  \end{subfigure} \hspace*{-0.45em}
  \begin{subfigure}[b]{0.12\textwidth}
    \centering
      \includegraphics[width=\textwidth, interpolate=false]{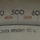}
  \end{subfigure} \hspace*{-0.45em}
  \begin{subfigure}[b]{0.12\textwidth}
    \centering
      \includegraphics[width=\textwidth, interpolate=false]{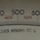}
  \end{subfigure}
  \hspace*{0.2em}
  \rotatebox[origin=l]{90}{\makebox[0.1\textwidth]{$\scriptstyle \times 2.5$}}
  \vspace*{-0.15em}

  \hspace*{-0.5em}
  \begin{subfigure}[b]{0.12\textwidth}
    \centering
      \includegraphics[width=\textwidth, interpolate=false]{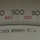}
      \caption{GT}
  \end{subfigure} \hspace*{-0.45em}
  \begin{subfigure}[b]{0.12\textwidth}
    \centering
      \includegraphics[width=\textwidth, interpolate=false]{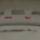}
      \caption{\textbf{RCAN}}
  \end{subfigure} \hspace*{-0.45em}
  \begin{subfigure}[b]{0.12\textwidth}
    \centering
      \includegraphics[width=\textwidth, interpolate=false]{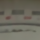}
      \caption{LIIF}
  \end{subfigure} \hspace*{-0.45em}
  \begin{subfigure}[b]{0.12\textwidth}
    \centering
      \includegraphics[width=\textwidth, interpolate=false]{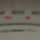}
      \caption{ArbSR}
  \end{subfigure} \hspace*{-0.45em}
  \begin{subfigure}[b]{0.12\textwidth}
    \centering
      \includegraphics[width=\textwidth, interpolate=false]{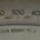}
      \caption{\textbf{IRN}}
  \end{subfigure} \hspace*{-0.45em}
  \begin{subfigure}[b]{0.12\textwidth}
    \centering
      \includegraphics[width=\textwidth, interpolate=false]{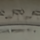}
      \caption{BAIRNet}
  \end{subfigure} \hspace*{-0.45em}
  \begin{subfigure}[b]{0.12\textwidth}
    \centering
      \includegraphics[width=\textwidth, interpolate=false]{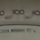}
      \caption{BAIRNet}
  \end{subfigure} \hspace*{-0.45em}
  \begin{subfigure}[b]{0.12\textwidth}
    \centering
      \includegraphics[width=\textwidth, interpolate=false]{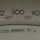}
      \caption{IARN}
  \end{subfigure}
  \hspace*{0.2em}
  \rotatebox[origin=l]{90}{\makebox[0.15\textwidth]{$\scriptstyle \times 3.5$}}

 \end{center}
 \vspace{-12pt}
    \caption{Visual examples of arbitrary rescaling at two scales: $\times 2.5$ and $\times 3.5$ (Best when zoomed-in to see better sharpness in IARN results).}
 \vspace{-12pt}
\label{fig:img}
\end{figure*}

\begin{figure*}[t]
\captionsetup[subfigure]{font=footnotesize, labelformat=empty}
\begin{center}

  \begin{subfigure}[b]{0.07\textwidth}
    \centering
      \includegraphics[width=\textwidth, interpolate=false]{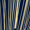}
  \end{subfigure} \hspace*{-0.5em}
  \begin{subfigure}[b]{0.07\textwidth}
    \centering
      \includegraphics[width=\textwidth, interpolate=false]{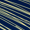}
  \end{subfigure} \hspace*{-0.5em}
  \begin{subfigure}[b]{0.07\textwidth}
    \centering
      \includegraphics[width=\textwidth, interpolate=false]{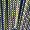}
  \end{subfigure} \hspace*{-0.5em}
  \begin{subfigure}[b]{0.07\textwidth}
    \centering
      \includegraphics[width=\textwidth, interpolate=false]{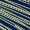}
  \end{subfigure} \hspace*{-0.5em}
  \begin{subfigure}[b]{0.07\textwidth}
    \centering
      \includegraphics[width=\textwidth, interpolate=false]{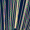}
  \end{subfigure} \hspace*{-0.5em}
  \begin{subfigure}[b]{0.07\textwidth}
    \centering
      \includegraphics[width=\textwidth, interpolate=false]{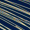}
  \end{subfigure} \hspace*{-0.5em}
  \begin{subfigure}[b]{0.07\textwidth}
    \centering
      \includegraphics[width=\textwidth, interpolate=false]{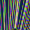}
  \end{subfigure} \hspace*{-0.5em}
  \begin{subfigure}[b]{0.07\textwidth}
    \centering
      \includegraphics[width=\textwidth, interpolate=false]{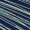}
  \end{subfigure} \hspace*{-0.5em}
  \begin{subfigure}[b]{0.07\textwidth}
    \centering
      \includegraphics[width=\textwidth, interpolate=false]{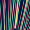}
  \end{subfigure} \hspace*{-0.5em}
  \begin{subfigure}[b]{0.07\textwidth}
    \centering
      \includegraphics[width=\textwidth, interpolate=false]{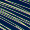}
  \end{subfigure} \hspace*{-0.5em}
  \begin{subfigure}[b]{0.07\textwidth}
    \centering
      \includegraphics[width=\textwidth, interpolate=false]{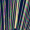}
  \end{subfigure} \hspace*{-0.5em}
  \begin{subfigure}[b]{0.07\textwidth}
    \centering
      \includegraphics[width=\textwidth, interpolate=false]{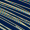}
  \end{subfigure} \hspace*{-0.5em}
  \begin{subfigure}[b]{0.07\textwidth}
    \centering
      \includegraphics[width=\textwidth, interpolate=false]{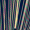}
  \end{subfigure} \hspace*{-0.5em}
  \begin{subfigure}[b]{0.07\textwidth}
    \centering
      \includegraphics[width=\textwidth, interpolate=false]{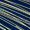}
  \end{subfigure}
  
  \begin{subfigure}[b]{0.14\textwidth}
    \centering
      \includegraphics[width=\textwidth, interpolate=false]{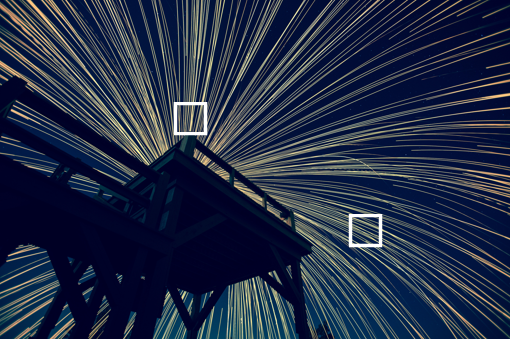}
      \caption{Bicubic}
  \end{subfigure} \hspace*{-0.5em}
  \begin{subfigure}[b]{0.14\textwidth}
    \centering
      \includegraphics[width=\textwidth, interpolate=false]{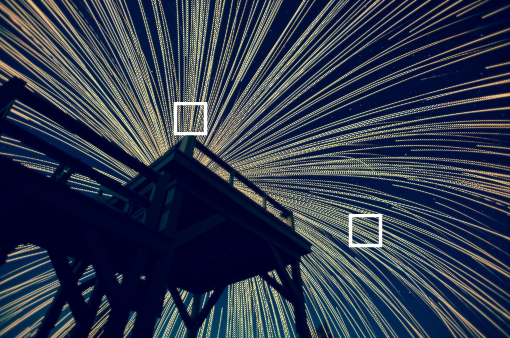}
      \caption{\textbf{CAR}}
  \end{subfigure} \hspace*{-0.5em}
  \begin{subfigure}[b]{0.14\textwidth}
    \centering
      \includegraphics[width=\textwidth, interpolate=false]{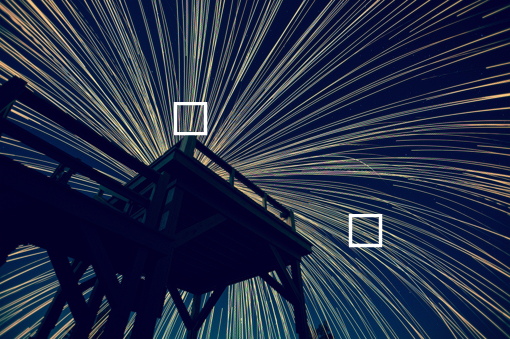}
      \caption{\textbf{IRN}}
  \end{subfigure} \hspace*{-0.5em}
  \begin{subfigure}[b]{0.14\textwidth}
    \centering
      \includegraphics[width=\textwidth, interpolate=false]{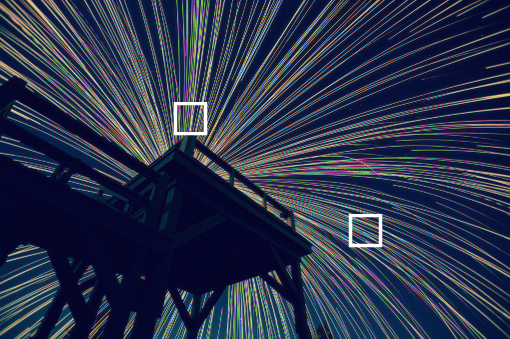}
      \caption{\textbf{HCFlow}}
  \end{subfigure} \hspace*{-0.5em}
  \begin{subfigure}[b]{0.14\textwidth}
    \centering
      \includegraphics[width=\textwidth, interpolate=false]{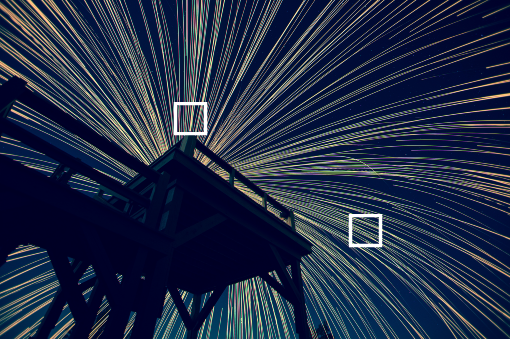}
      \caption{BAIRNet}
  \end{subfigure} \hspace*{-0.5em}
  \begin{subfigure}[b]{0.14\textwidth}
    \centering
      \includegraphics[width=\textwidth, interpolate=false]{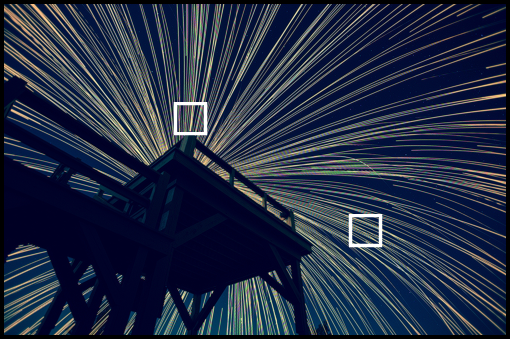}
      \caption{AIDN}
  \end{subfigure} \hspace*{-0.5em}
  \begin{subfigure}[b]{0.14\textwidth}
    \centering
      \includegraphics[width=\textwidth, interpolate=false]{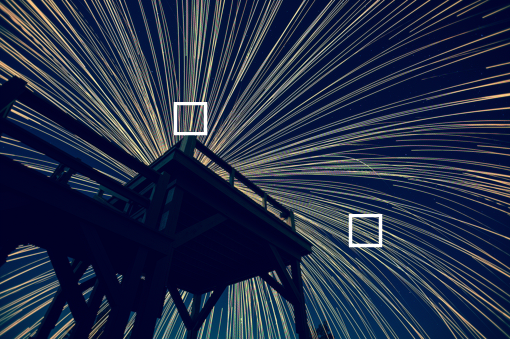}
      \caption{IARN}
  \end{subfigure}

 \end{center}
   \vspace{-12pt}
    \caption{Visual examples of generated LR images with magnified view of worst-case visual artifacts .}
   \vspace{-3pt}
\label{fig:lr}
\end{figure*}

As shown in Table~\ref{tab:main}, results from our IARN are the best for all test cases except for a couple of
cases where AIDN is slightly better.
Using the challenging Manga109 test set as an example, it leads over BAIRNet in PSNR
by $+3.04$, $+1.16$ and $+0.89$ for $\times1.5$, $\times2.5$ and $\times3.5$ respectively.
As shown in visual examples in Fig.~\ref{fig:img}, bidirectional methods like IRN, BAIRNet, AIDN and our IARN
outperform others significantly.  Among the four, IARN is able to restore details more accurately, making it
easier to recognize details like the number $\mathbf{00}$.
The IARN model used for comparisons here uses an asymmetric pre-training for the first stage.
For intra-model comparison, two other variants, IARN$^\dagger$ and IARN$^\ddagger$,
are also included in Table~\ref{tab:main} with $\times 3.5$ as an example.
IARN$^\dagger$ takes the same pre-training strategy
but only uses DIV2K with baseline batch and patch sizes, resulting a slightly worse performance behind IARN.
IARN$^\ddagger$ is the same as IARN$^\dagger$ in data settings but it is exclusively trained using
symmetric scales for both stages.  It is consistently worse than IARN, proving that pre-training with asymmetric scales
can help the final model performs better in tests of symmetric scales.

\begin{figure}[t!]
\begin{center}
  \begin{subfigure}[b]{0.48\textwidth}
    \centering
      \includegraphics[width=\textwidth]{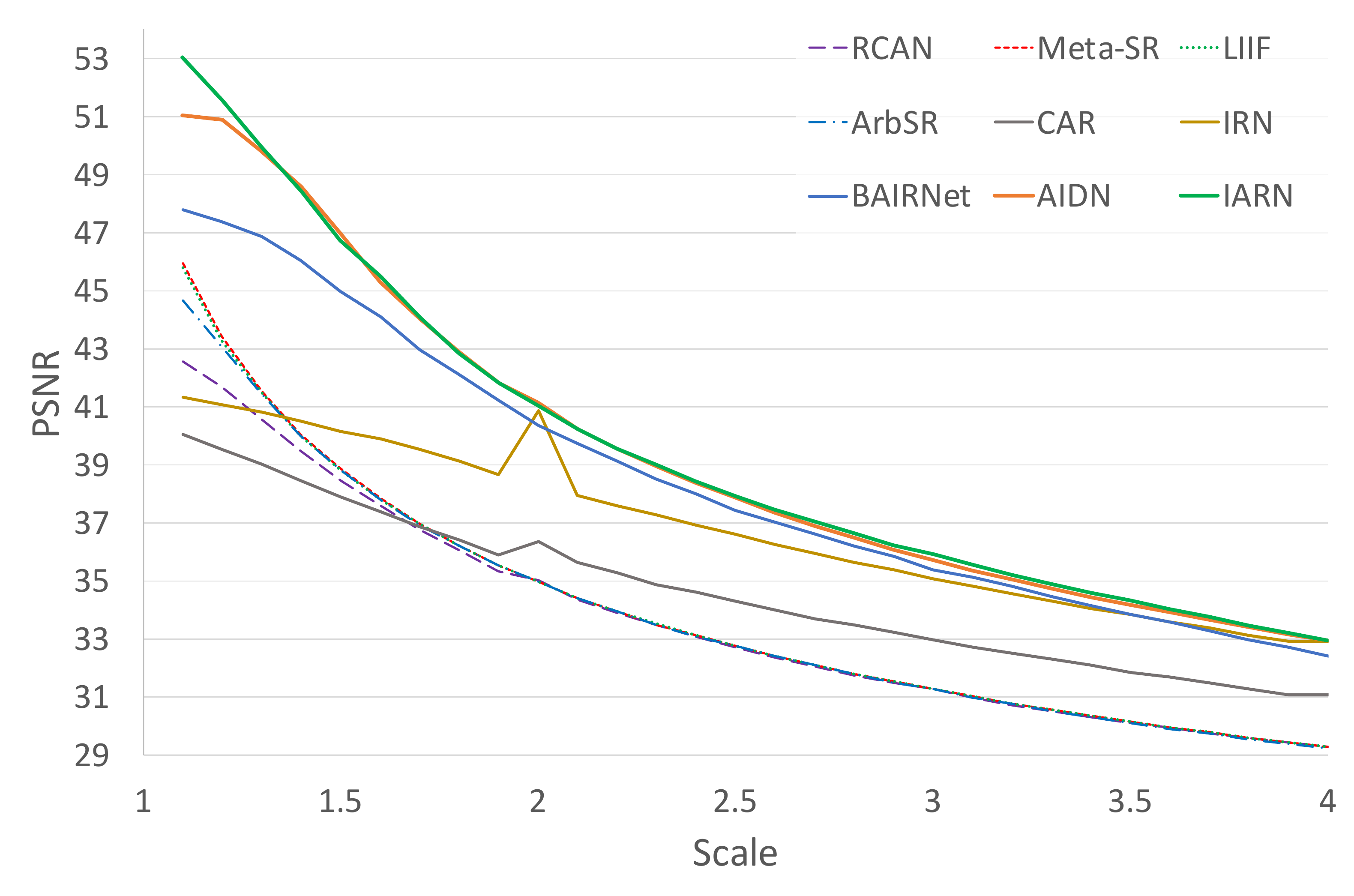}
  \end{subfigure}
\end{center}
\vspace{-10pt}
 \caption{Cross-scale ($\times 1.1-\times 4$) performance for arbitrary rescaling testing of DIV2K validation set.}
 \vspace{-12pt}
 \label{fig:scale}
 \end{figure}

\footnotetext{Incomplete AIDN results due to lack of model or data.}

Results of continuous scales between $\times 1.1$ and $\times 4$ (increasing by 0.1 at each sampling point)
are also illustrated in Fig.~\ref{fig:scale} for a more thorough assessment.
As explained in BAIRNet, there is a significant performance boost when the models are trained jointly for both
downscaling and upscaling.  Among this group, BAIRNet is obviously better than CAR
and IRN across the range overall, except trailing slightly behind
IRN for a limited choices of scales.  Both AIDN and our IARN model are able to outperform other models across
the full range of arbitrary scales, surpassing IRN even at fixed scales like $\times 2$ and $\times 4$.
Between the two, our IARN has better performance for most of the scales, especially for small scales close to
$\times 1.1$ and large scales around $\times 3$.

\subsection{LR Image Quality}

For the group of six bidirectional rescaling methods, LR outputs are also generated from the learning based models respectively.
Without a known ground-truth reference, the goal is to generate visually pleasant images which look similar to
the LR reference that is downsampled from the HR input using conventional bicubic interpolation.
As shown in Table~\ref{tab:lr}, the SSIM values calculated in related to the LR references and the blind image quality
metric NIQE~\cite{mittal_spl_2012} are compared
for two scales, $\times 2$ and $\times 4$,
to accommodate the native integer scales for CAR and IRN.
It is clearly shown that IRN and IARN are the best two, while AIDN is trailing behind as the third best.
Visual examples of $\times 4$ LR images with the most noticeable false color artifacts are included in Fig.~\ref{fig:lr}
to compare different models.
While they are not much different from the bicubic reference in overall view, the CAR
one is noticeably brighter.  Among others, HCFlow and BAIRNet are obviously the
worst two.  While the remaining three are close,
false-color artifacts are more noticeable in AIDN as seen from the left zoomed-in window.

 \begin{table}[h!]
	\centering
	\footnotesize
	\setlength{\tabcolsep}{1.5pt}
\vspace{0pt}
	\caption{Quantitative quality assessment (NIQE$\downarrow$/SSIM$\uparrow$) of generated LR outputs.} 
\vspace{3pt}
	\begin{tabular}{ccccc} 
		\hline
		{Method} & {S} & {Urban100$\wt{^{^a}}$} & {Manga109$\wt{^{^a}}$} & {DIV2K$\wt{^{^a}}$} \\
		\hline \hline
		CAR~\cite{sun_tip_2020}$\wt{^{^a}}$ & {2} & 6.360/0.9658 & 4.365/0.9772 & 3.660/0.9748\\
		IRN~\cite{xiao_eccv_2020}& {2} & \blue{5.956}/\blue{0.9941} & \blue{4.214}/\blue{0.9959} & \red{3.347}/\blue{0.9945} \\
	    BAIRNet~\cite{pan_cvpr_2022} & {2} & 6.475/0.9797 & 4.431/0.9892 & 3.507/0.9864\\
	    AIDN~\cite{xing_arxiv_2022} & {2} & -\footnotemark[\value{footnote}] & -\footnotemark[\value{footnote}] & 3.516/0.9920\\
	    {IARN}$\wt{A}$ \vspace{1pt} & {2} & \red{5.783}/\red{0.9963} & \red{4.120}/\red{0.9971} & \blue{3.359}/\red{0.9963} \\
		\hline
		CAR~\cite{sun_tip_2020}$\wt{^{^a}}$ & 4 & 22.731/0.9196 & 6.886/0.9529 & 5.549/0.9460\\
		IRN~\cite{xiao_eccv_2020}& {4} & \blue{18.035}/\red{0.9916} & \red{5.884}/\blue{0.9932} & \blue{4.094}/\blue{0.9933} \\
		HCFlow~\cite{liang_iccv_2021} & {4} & 18.475/0.9651 & 6.457/0.9738 & 4.744/0.9785\\
	    BAIRNet~\cite{pan_cvpr_2022} & {4} & 19.401/0.9716 & 7.632/0.9824 & 4.896/0.9841\\
	    AIDN~\cite{xing_arxiv_2022} & {4} & -\footnotemark[\value{footnote}] & -\footnotemark[\value{footnote}] & 4.165/0.9909\\
	    {IARN}$\wt{A}$ \vspace{1pt} & {4} & \red{18.020}/\red{0.9928} & \blue{5.961}/\red{0.9932} & \red{4.088}/\red{0.9944} \\
		\hline	\end{tabular}
\label{tab:lr}
\vspace{-8pt}
\end{table}

\begin{table}[h!]
	\centering
	\footnotesize
	\setlength{\tabcolsep}{1pt}
\vspace{0pt}
	\caption{Ablation study for different modules (Color-highlighted group in one column means they share the same setting in all other columns).}
\vspace{3pt}
	\begin{tabular}{cccccc} 
		\hline
		{Channel} & {Scale} & {Atrous} & {Rescaling} & {\#} & {PSNR$\wt{^{^a}}$} \\
		{Splitting} & {Encoding} & {Convolution} & {Method} & {InvBlock} & {Urban100 $\times 4$} \\
		\hline \hline
		{\red{\xmark}} & {Dual} & {\xmark} & Bilinear & 16 & 26.79 \\
		{\red{\cmark}} & \blue{Dual} & \green{\xmark} & Bilinear & 16 & 27.22 \\
		{\cmark} & \blue{HF} & {\xmark} & Bilinear & 16 & 26.93 \\
		{\cmark} & \blue{LF} & {\xmark} & Bilinear & 16 & 27.04 \\
		{\cmark} & \blue{None} & {\xmark} & Bilinear & 16 & 26.70 \\
		{\cmark} & {Dual} & \green{\cmark} & \teal{Bilinear} & 16 & 27.41 \\
		{\cmark} & {Dual} & {\cmark} & \teal{NN} & \orange{16} & 27.50 \\
		{\cmark} & {Dual} & {\cmark} & \teal{Bicubic} & 16 & 27.11 \\
		{\cmark} & {Dual} & {\cmark} & NN & \orange{20} & 27.76 \\
		{\cmark} & {Dual} & {\cmark} & NN & \orange{24} & 27.79 \\
		\hline	\end{tabular}
\label{tab:ablation}
\vspace{-1pt}
\end{table}

\begin{figure*}[t!]
\captionsetup[subfigure]{font=footnotesize, labelformat=empty}
\begin{center}
  \hspace*{2em}
  \begin{subfigure}[b]{0.105\textwidth}
    \centering
      \includegraphics[width=\textwidth, interpolate=false]{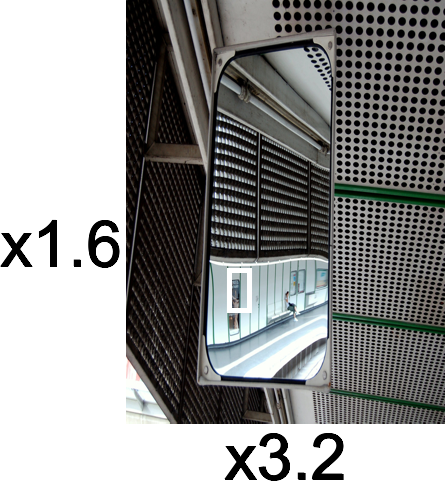}
      \caption{LR\wt{$^{\star}$}}
  \end{subfigure} \hspace*{1.03em}
  \begin{subfigure}[b]{0.12\textwidth}
    \centering
      \includegraphics[width=\textwidth, interpolate=false]{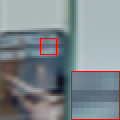}
      \caption{Bicubic\wt{$^{\star}$}}
  \end{subfigure} \hspace*{-0.4em}
  \begin{subfigure}[b]{0.12\textwidth}
    \centering
      \includegraphics[width=\textwidth, interpolate=false]{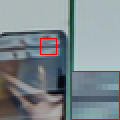}
      \caption{ArbSR\wt{$^{\star}$}}
  \end{subfigure} \hspace*{-0.4em}
  \begin{subfigure}[b]{0.12\textwidth}
    \centering
      \includegraphics[width=\textwidth, interpolate=false]{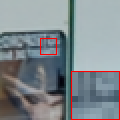}
      \caption{\textbf{IRN}\wt{$^{\star}$}}
  \end{subfigure} \hspace*{-0.4em}
  \begin{subfigure}[b]{0.12\textwidth}
    \centering
      \includegraphics[width=\textwidth, interpolate=false]{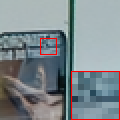}
      \caption{BAIRNet\wt{$^{\star}$}}
  \end{subfigure} \hspace*{-0.4em}
  \begin{subfigure}[b]{0.12\textwidth}
    \centering
      \includegraphics[width=\textwidth, interpolate=false]{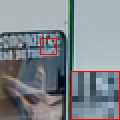}
      \caption{IARN$^\star$}
  \end{subfigure} \hspace*{-0.4em}
  \begin{subfigure}[b]{0.12\textwidth}
    \centering
      \includegraphics[width=\textwidth, interpolate=false]{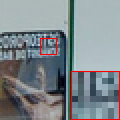}
      \caption{GT\wt{$^{\star}$}}
  \end{subfigure}

 \end{center}
\vspace{-12pt}
    \caption{Visual examples of arbitrary asymmetric scales.}
\label{fig:xy}
\vspace{-12pt}
\end{figure*}

\subsection{Ablation Study}

To show the effects of different modules proposed here, we conduct a comprehensive ablation study for
different combinations using Urban100 $\times 4$ as the testing benchmark.
For illustrative purpose, all models were trained for $100k$ iterations and the differences
in performance is clearly shown in Table~\ref{tab:ablation}.
For easy cross-reference, colored highlight are used to associate mini-groups for comparison.
Two of the most critical modules to boost performance are preemptive channel splitting and
position-aware scale encoding.  As highlighted in \red{red}, when channel-splitting is activated while keeping other
configurations identical, the average PSNR increases by $+0.43$.  Note that when no channel-splitting is applied, original
RGB channels are fed to the LF branch while inputs to HR are zeros.
For scale encoding options as highlighted in \blue{blue}, it is the best to include it for both branches, boosting PSNR by $+0.52$.
When atrous convolution is applied, another $+0.19$ is added in PSNR.  For rescaling, NN is the best choice while bicubic is far
behind the other two.  This corroborates the methodology discussed earlier that channel-split function $s(\cdot)$ makes
$\mathbf{x}^{lf}_H$ rescaling-invertible when using NN resampling.
While the final PSNR is always better when
the number of InvBlock layers increases, it is set at 20 in other experiments for better efficiency with minimum performance loss.

 \begin{table}[h]
 \footnotesize
 \setlength{\tabcolsep}{2pt}
     \vspace{-3pt}
     \caption{PSNR for symmetric and asymmetric scale factors.}
     \vspace{-3pt}
 \begin{center}
 \begin{tabular}{ccccccccccccc}
 \hline
 \multirow{2}{*}{} & \multicolumn{4}{c}{BSD100} & \multicolumn{4}{c}{Urban100} & \multicolumn{4}{c}{$^{\wt{\ddagger}}$Manga109}\\
 & $\frac{\times 2}{\times 3}$ & $\frac{\times 1.6}{\times 3.2}$ & $\frac{\times 3.6}{\times 1.2}$ & $\times 2.5$
 & $\frac{\times 2}{\times 3}$ & $\frac{\times 1.6}{\times 3.2}$ & $\frac{\times 3.6}{\times 1.2}$ & $\times 2.5$
 & $\frac{\times 2}{\times 3}$ & $\frac{\times 1.6}{\times 3.2}$ & $\frac{\times 3.6}{\times 1.2}$ & $\times 2.5$ \vspace{1pt}\\
 \hline \hline
  ArbSR~\cite{wang_iccv_2021}$^{\wt{\ddagger}}$ & 30.58 & 30.87 & 30.24 & 30.51 & 30.59 & 30.60 & 29.74 & 30.68 & 36.17 & 35.88 & 35.30 & 36.67 \\
  \textbf{IRN}~\cite{xiao_eccv_2020} & 34.87 & 34.69 & 34.19 & 35.56 & 32.75 & 32.44 & 32.09 & 33.99 & 37.42 & 37.08 & 37.12 & 39.33 \\
   BAIRNet~\cite{pan_cvpr_2022} & \blue{36.28} & \blue{36.96} & \blue{36.64} & \blue{36.17} & \blue{36.32} & \blue{36.74} & \blue{35.82} & \blue{36.43} & \blue{40.01} & \blue{40.27} & \blue{39.42} & \blue{40.14} \\
   IARN$^\star$ & \red{36.96} & \red{38.04} & \red{39.17} & \red{36.65} & \red{37.06} & \red{37.92} & \red{38.56} & \red{36.83} & \red{41.59} & \red{42.12} & \red{42.32} & \red{41.38} \\
 \hline
 \end{tabular}
 \end{center}
 \label{tab:xy}
 \vspace{-12pt}
 \end{table}

\subsection{Asymmetric Arbitrary Rescaling}
As discussed earlier, our IARN model can be optimized for asymmetric scales using the exact same network architecture.
For fair comparisons with ArbSR and BAIRNet,
another model denoted as IARN$^\star$ is trained using asymmetric scale in both stages.
As shown in Table~\ref{tab:xy}, three large benchmark test
sets are used for assessment in three asymmetric and one symmetric scales.
Results from ArbSR, IRN and BAIRNet are used for comparisons, using the same evaluation protocol
as in BAIRNet.  It is demonstrated that, both in quantitative metrics and visual
exampled as included in Fig.~\ref{fig:xy}, our IARN$^\star$ holds a clear advantage
over other peers in restoring more accurate details.

\subsection{Model Complexity and Efficiency}
\vspace{-2pt}

To assess the efficiency of our proposed IARN, it is compared with others in terms of model size and average inference time per image for the BSD100 test set.  As illustrated in Table~\ref{tab:efficency},
three most relevant models which are optimized jointly for downscaling and upscaling are included so the inference time includes both downscaling and upscaling and listed separately.
Note that for CAR and IRN, they have models of different sizes depending on the scale factor, while for BAIRNet and our IARN, only one
model is needed for both tests.  It is shown that our IARN is comparable to IRN $\times 4$ in terms of number of parameters, much more smaller
than CAR and BAIRNet.  For inference speed, it is slower than CAR and IRN.
But for the only two models capable of bidirectional arbitrary rescaling (AIDN not yet available for assessment),
our model is able to outperform the previous SOTA BAIRNet consistently while using less than 20\% parameters and reducing
inference time by 66\% and 60\% for $\times 2$ and $\times 4$ respectively.

\vspace{-3pt}
\begin{table}[h!]
	\centering
	\footnotesize
	\setlength{\tabcolsep}{2pt}
\vspace{0pt}
	\caption{Model size and inference time comparison.}
\vspace{0pt}
	\begin{tabular}{cccccccccc} 
		\hline
		{Method$\wt{^{^a}}$} & {Scale} & {Param} & {Downscaling} & {Upscaling} & {Method$\wt{^{^a}}$} & {Scale} & {Param} & {Downscaling} & {Upscaling} \\
		\hline \hline
		\multirow{2}{*}{CAR~\cite{sun_tip_2020}} & {$\times 2$} & 51.1M & 0.004$s$ & 0.005$s$ & \multirow{2}{*}{IRN~\cite{xiao_eccv_2020}}$\wt{^{^a}}$ & {$\times 2$} & 1.66M & 0.018$s$ & 0.021$s$ \\
		  & {$\times 4$} & 52.8M & 0.004$s$ & 0.005$s$ & & {$\times 4$} & 4.35M & 0.025$s$ & 0.026$s$ \\
		\hline
	    \multirow{2}{*}{BAIRNet~\cite{pan_cvpr_2022}} & {$\times 2$} & \multirow{2}{*}{22.4M} & 0.506$s$ & 0.129$s$ & \multirow{2}{*}{IARN} $\wt{^{^a}}$ & {$\times 2$} & \multirow{2}{*}{4.32M} & 0.058$s$ & 0.154$s$ \\
	     & {$\times 4$} & & 0.459$s$ & 0.061$s$ & & {$\times 4$} & & 0.058$s$ & 0.154$s$ \\
		\hline	\end{tabular}
\label{tab:efficency}
\vspace{-12pt}
\end{table}

\section{Conclusions}

Image arbitrary rescaling using deep learning is a relatively new and under explored topic in the field of low-level image processing due to its complexity.
In this paper, we have presented the first invertible arbitrary image rescaling work.
Based on an INN backbone enhanced with a novel preemptive channel splitting module and a new position-aware scale encoding method,
the newly proposed IARN network is capable of handling bidirectional image arbitrary rescaling over a large range of scales using just one trained model.
Extensive experiments on a comprehensive set of benchmark datasets validate a much better performance of arbitrary image rescaling over
the current related SOTA methods in both HR and LR outputs, with reduced model size and faster inference comparing to BAIRNet.
In addition, better performance is also achieved for asymmetric arbitrary rescaling tests.

\bibliographystyle{unsrt}

\end{document}